\documentclass[12pt]{article}
\usepackage{amsmath}
\usepackage{amsthm}
\usepackage{amssymb}
\usepackage{color}
\usepackage{amsfonts}
\usepackage{epic}
\usepackage{eepic}
 \usepackage{times}
\usepackage{graphics}
\usepackage[matrix,arrow]{xy}
\usepackage{macros}
\usepackage{epsfig}
\usepackage{floatflt}
\usepackage[centerlast]{caption2}
\def\Between#1#2#3#4{ %% #1#2 : size, #3 : place, #4 : letter
\raise-#1mm\vbox to#1mm{\hsize #2mm \vbox{\vskip #3mm\centerline{#4}
} } }
\def\Dummy{{{ }\over{ }}}
\def\generalYoung{
\Between{10}{15}7{$\lambda=$}
 \Square{20}9{$r_{n-1}$}{$s_{n-1}\Dummy$} \kern-.4pt
 \Square{17}8{$r_{n-2}$}{$s_{n-2}\Dummy$} \kern-.4pt
\Between{10}{15}7{$\cdots\cdots$}
 \Square{13}6{$r_2    $}{$\;s_2  \Dummy$} \kern-.4pt
 \Square{10}5{$r_1    $}{$\;s_1  \Dummy$} \kern-.4pt
\Between{15}{10}{10}{\hfill .} }
\def\generalYoung{
\vskip.25cm \noindent \makebox[  4cm]{ } \makebox[
2cm]{$s^1$}\hskip-.4pt \makebox[1.7cm]{$s^2$} \makebox[1.4cm]{ }
\makebox[1.4cm]{$s^{N-2}$}\hskip-.35pt \makebox[1.3cm]{$s^{N-1}$}
\hfill\break
 \makebox[  4cm][r]{$\hfill\lambda=$}
\framebox[  2cm][l]{\rule[  -1cm]{0cm}{  2cm}$r^1$}\hskip-.4pt
\framebox[1.7cm][l]{\rule[-0.7cm]{0cm}{1.7cm}$r^2$}
 \makebox[1.4cm]                {\raisebox{.25cm}{$\cdots\cdots$}}
\framebox[1.4cm][l]{\rule[-0.4cm]{0cm}{1.4cm}\raisebox{.25cm}{$r^{N-2}$}
                                 }\hskip-.4pt
\framebox[1.3cm][l]{\rule[-0.2cm]{0cm}{1.2cm}\raisebox{.25cm}{$r^{N-1}$}}
\makebox[1cm][r]{.} \vskip.3cm }
\theoremstyle{plain}

\theoremstyle{remark}

\newcommand{\beq}{\begin{equation}}
\newcommand{\eeq}{\end{equation}}

\newcommand{\ket}[1]{|#1\rangle}
\newcommand{\bra}[1]{\langle#1|}

\newcommand{\aufz}
{\begin{list}{$\bullet$}{\topsep0cm \itemsep0cm \parsep0cm}}
\newcommand{\eaufz}{\end{list}}

\def \ignore#1 { {} }

\def \Fig#1#2#3 {
\begin{figure}
\begin{center}
\scalebox{.6}{\includegraphics{#1.eps}} \label{#1}
\end{center}
\caption{#3}
\end{figure}
}

\def \p {\partial}
\def \pp#1 {{\frac{\p}{\p #1}}}
\def \ppd#1 {{\frac{\p^2}{\p #1 ^2}} }

\def \vir { \ , \ }
\def \F32#1#2#3#4#5#6{{} _3F_2\left(\left.\begin{array}{c}#1 \vir #2 \vir
    #3 \\ #4\vir #5
    \end{array} \right| #6 \right) }

\def \bea {\begin{eqnarray}}
\def \eea {\end{eqnarray}}
\def \bee {\begin{eqnarray*}}
\def \eee {\end{eqnarray*}}

\begin{document}

\title{Note on refined topological vertex, Jack symmetric functions and instanton counting (I)}

\author{Jian-feng Wu$^{(1)}$}

\address{
$^{(1)}$ Institute of Theoretical Physics, Chinese Academy of
Sciences, Beijing, China}

\maketitle

\section*{Abstract}

In this article, we calculated the refined topological vertex for
the one parameter case using the Jack symmetric functions. Also, we
obtain the partition function for elliptic N=2 models, the results
coincide with those of Nekrasov instanton counting partition
functions for the $N=2^{\ast}$ theories.

\section{Introduction}

The study of refined topological vertex sheds lights on many other
physicial or mathematical problems in recent years. For physical
interests, we are interested in the $N=2$ gauge system[1-10], which
can be realized as a IIA string theory compactifying on certain
toric Calabi-Yau three-folds(CY-3folds) by the geometric
engineering. The instanton part of the N=2 theory is captured by the
topological string amplitude. This amplitude, can be calculated by
the refined topological vertex\cite{TopVertex} formulation.
Otherwise, the $N=2$ systems can also have the standard NS5-D4 brane
configurations and show a great many interesting properties, such as
the S-duality, the natural confinement, the integrability and so on
and so forth. Recently, Alday, Gaitto and Tachikawa(AGT)\cite{AGT}
showed that for an arbitrary $N=2$ $SU(2)$ superconformal gauge
system, there is a dual 2d theory which is a Liouville theory living
on the moduli space of the 4d gauge theory[13-31]. The moduli space
(Seiberg-Witten curve) of the $N=2$ theory is also the ramification
of the Riemann surface M5 branes wrapping on. Later on, Dijkgraaf
and Vafa\cite{DV2009a} proved this 2d-4d relation using intrinsic
correspondences of topological string-matrix model-Liouville theory.
Recently, Cheng, Dijkgraaf and Vafa\cite{DV2010b} extended this
proof to more detailed cases, they showed that the instanton part of
Nekrasov partition function, which duals to the conformal block of
the 2d conformal field theory(CFT), is in fact a linear combination
of the non-perturbative string partition functions.

The NS5-D4 brane configuration of a given $N=2$ gauge theory has a
rather simple translation to topological string. Roughly speaking,
the brane configuration diagram can be seen as the singular version
of the related toric diagram of the related CY-3fold. For example,
if we consider there are N coincided D4 branes truncated by 2
separated NS5 branes, which low energy theory is a 4d $N=2$ $U(N)$
gauge theory with $N_f = 2N$ fundamental matters. Classically, there
are two N-fold singularities located at the two intersection points
of the branes. However, while lift this to M-theory, these
singularities can be "blown-up" to an sequence of $S^{2}$s due to
the quantum effect. Then the brane diagram becomes an N-ramified
Riemann sphere on which M5 branes wrapping on. On the topological
string side, to get the same gauge theory, by the standard geometric
engineering, one can identify the "blow-up" process that the
conifold singularities transit to resoloved ones. Fig.1 shows the
simple case for $N=2$ $U(2)$ gauge theory.

\begin{figure}
\centering
\includegraphics{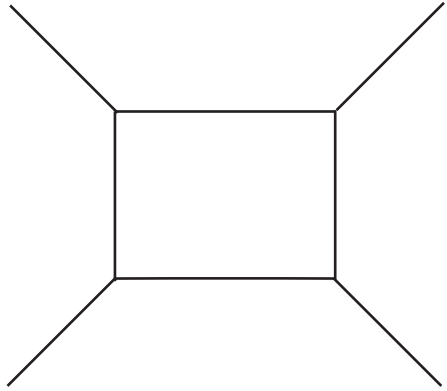}
\caption{\small{\emph{Toric diagram of U(2) pure gauge theory}}}
\label{Fig1.eps}
\end{figure}

The topological string partition function on a given toric CY3-fold
is expected to correspond to instanton sums in the related gauge
theory. The instanton part of the partition function of a certain
$N=2$ gauge theory can also have a brane expression, the D0-D4
configuration. In this configuration, the D0 branes dissolve into
the D4 branes as the instanton background of the gauge theory. These
D0 branes come from the M-theory compactification as the
Kaluza-Klein modes. Apart from M5 branes, there are M2 branes which
are the magnetic dual of the M5 branes. Then if there are M2 branes
intersecting with M5 branes, then after the M-theory
compactification, they become the D2-D0 bound states. If these D2
branes wrap on some nontrivial Lagrangian 2-cycles in the CY3-fold,
they behave just like D0 branes for the observer living on the D4
branes, thus they also contribute to the instanton counting of the
$N=2$ gauge theories. So the instantons of the $N=2$ theory are
expected to relate to the D2-D0 bound states. Besides, as noted in
\cite{RTV}, the instanton calculation has more refined information.
This information comes from the Nekrasov $\Omega$ deformation of
$N=2$ gauge theories. From the M-theory point of view, the CY3
compactification gives a 5d gauge theory living on
$\mathbb{C}^{2}\times S^1$ with $S^{1}$ the M-theory cycle. The BPS
spectrum of the theory corresponds to the little group
representation of the motion group of $\mathbb{C}^{2}\times S^1$,
which is, obviously, the $SO(4)=SU(2)_L \times SU(2)_R \in SO(5)$.
The $\Omega$ deformation is a $T^2$ action on $\mathbb{C}^2$:
\begin{equation}
T^2 : (z_1,
z_2)\longmapsto\,\,(e^{i\epsilon_1}z_2,e^{i\epsilon_2}z_2).
\end{equation}
This deformation has a direct impact on the definition of the
topological amplitude, which can be easily calculated by the
topological vertex formulism. Now the fundamental vertices change to
the refined ones which are the two-parameter generalization of the
original topological vertices. This change is due to the fact that
topological string amplitude counts the holomorphic maps from string
world sheet to Langrangian submanifolds of the toric CY3-fold. On
the other hand, the maps also correspond to BPS bound
states\cite{Rajesh-Vafa1, Rajesh-Vafa2} of M2 branes, which are
representations of $SO(4)=SU(2)_L \times SU(2)_R$, now twisted by
the $\Omega$ deformation. Thus the refined topological vertex is a
fundamental block for building the $N=2$ theories.

\begin{figure}
\centering
\includegraphics{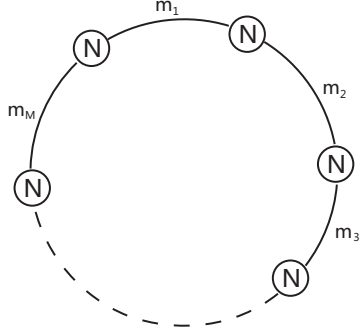}
\caption{\small{\emph{M-node necklace quiver U(N) gauge theory}}}
\label{Fig2.eps}
\end{figure}

It is crucial that this observation also implies the AGT relation,
and reveals the essential net dualities between these topological
string-matrix model and $N=2$ 4d gauge theories-2d conformal field
theories. However, for generic $N=2$ theories, it is hard to verify
these dualities, since there are many ambiguities in all of these
theories. The simpler cases are the so called "$N=2^{\ast}$ theory",
which only involves an adjoint matter, and the "necklace" quiver
$N=2$ theories which only have bifundamental matters. \footnote{For
convenience, we call all the elliptic $N=2$ models $N=2^{\ast}$
theories.}In this article, we will concentrate on these theories.
Their brane configurations are just the elliptic models, which have
punctured torus $\mathcal{T}_{M,1}$ as Seiberg-Witten curves. The
related integrable system is the two dimensional elliptic
Calogero-Sutherland(eCS) model, from which one can easily read off
the Liouville/Toda theories living on torus. From either the
topological string or the eCS model, one can have a rather simple
description of the one parameter refined topological vertex and
further the instanton counting by invoking the Jack symmetric
functions. We found that the refined topological vertex have a
simple description by using the Jack polynomials. The instanton
counting of $N=2$ quiver theories can also be computed by the same
Jack polynomials. The main result of this article is the following
closed formulae for $N=2$ $U(N)$ $M$-node($M\geq2$) necklace quiver
gauge theories(see Fig.\ref{Fig2.eps}) \vspace{10mm}

\begin{eqnarray}
&&{\bf
Z}_{bifund}^{4D\,\,\,inst}(\vec{a}_{\ell},\vec{\lambda}_{\ell},\vec{a}_{\ell
+1},\vec{\lambda}_{\ell +1}; m_{\ell}) = \prod_{m,n=1}^{N}\langle
E^{m^{(\ell,\ell+1)}_{m,n}}(E^{\ast})^{\beta -
m^{(\ell,\ell+1)}_{m,n}-1} J_{\lambda_{\ell,m}},
J_{\lambda_{\ell+1,n}}\rangle_{\beta} \\ &&{\bf
Z}_{adj}^{4D\,\,\,inst}(\vec{a}_{\ell},\vec{\lambda}_{\ell};
m_{\ell})={\bf
Z}_{bifund}^{4D\,\,\,inst}(\vec{a}_{\ell},\vec{\lambda}_{\ell},\vec{a}_{\ell},\vec{\lambda}_{\ell};
m_{\ell})\nonumber\\\nonumber &&{\bf
Z}_{vec}^{4D\,\,\,inst}(\vec{a},\vec{\lambda}) = 1/{\bf
Z}_{adj}^{4D\,\,\,inst}(\vec{a},\vec{\lambda};0)\\\label{M-Necklace}
&&{\bf
Z}^{U(N)\,\,\,inst}_{M-necklace}=\sum_{\vec{\lambda}_1,\cdots,\vec{\lambda}_M}\prod_{i,\ell=1}^{M}
\tilde{Q}_{i}^{|\vec{\lambda}_i|}{\bf
Z}_{vec}(\vec{a}_i,\vec{\lambda}_i){\bf
Z}_{bifund}(\vec{a}_{\ell},\vec{\lambda}_{\ell},\vec{a}_{\ell+1},\vec{\lambda}_{\ell+1};m_{\ell})\nonumber\\&&=\sum_{\vec{\lambda}_1,\cdots,\vec{\lambda}_M}\prod_{i,\ell=1}^{M}(
\tilde{Q}_{i})^{|\vec{\lambda}_i|}\prod_{j,k=1}^{N}{\Large\mbox
{$[$}}(\langle E^{a^{(i)}_{j,k}}(E^{\ast})^{\beta - a^{(i)}_{j,k}-1}
J_{\lambda_{i,j}}, J_{\lambda_{i,k}}\rangle_{\beta}|_{j\neq
k})\langle J_{\lambda_{i,j}},
J_{\lambda_{i,j}}\rangle_{\beta}{\Large\mbox
{$]$}}^{-1}\nonumber\\&&\times \prod_{m,n=1}^{N}\langle
E^{m^{(\ell,\ell+1)}_{m,n}}(E^{\ast})^{\beta -
m^{(\ell,\ell+1)}_{m,n}-1} J_{\lambda_{\ell,m}},
J_{\lambda_{\ell+1,n}}\rangle_{\beta}.
\end{eqnarray}
 Here $\vec{a}_{\ell}=\{a_{\ell,1},\cdots,a_{\ell,N}\}$
and
$\vec{\lambda}_{\ell}=\{\lambda_{\ell,1},\cdots,\lambda_{\ell,N}\}$
defines the Coulomb parameter vector and instanton partition Young
tableau vector of the $\ell$-th $U(N)$ gauge group, respectively.
$m_{\ell}=\frac{\tilde{m}}{\epsilon_2}$ denotes the mass of the
$\ell$-th bifundamental matter.\[\tilde{Q}_i = \text{exp}(2\pi i
\tau_{UV}^i),\,\,\,\,\, \tau_{UV} = \frac{4\pi i}{g_{UV}^2} +
\frac{\theta_{UV}}{2\pi}\] are the sewing parameters.
$a^{(i)}_{j,k}=a_{i,j} - a_{i,k},
a_{i,j}=\tilde{a}_{i,j}/\epsilon_2,\,\, m^{(\ell,\ell+1)}_{m,n} =
a_{\ell+1,n}-a_{\ell,m}-m_{\ell}.$
\[E=1+e_{[1]}+e_{[2]}+\cdots=\text{exp}{\Large\mbox
{(}}\sum_{n>0}\frac{(-1)^n}{n}{p_n}{\Large\mbox {)}}\] is related to
Dijkgraaf-Vafa's topological B-brane background and this will be
shown explicitly in the second part of this note\cite{WuJF10b},
$e_{[m]}, p_n$ is elementary and power sum polynomials respectively.
$E^{\ast}$ is the adjoint action under the inner product of Jack
polynomials \cite{Okounkov}\[\langle E J_{\lambda},
J_{\mu}\rangle_{\beta}=\langle J_{\lambda}, (E^{\ast})
J_{\mu}\rangle_{\beta},\,\,\,\beta=-\frac{\epsilon_1}{\epsilon_2}.\]
The inner product is defined and proved in \cite{Okounkov} as
following
\begin{eqnarray}
\label{Okounkov}\langle
E^m(E^{\ast})^{\beta-m-1}J_{\lambda},J_{\mu}\rangle_{\beta}&=&(-1)^{\lambda}\beta^{-|\lambda|-|\mu|}\prod_{s\in
\lambda}(m+a_{\lambda}(s)+1+\beta
l_{\mu}(s))\\\nonumber&\times&\prod_{t\in
\mu}(m-a_{\mu}(t)-\beta(l_{\lambda}(t)+1)),
\end{eqnarray}
here \[a_{\lambda}(s)=\lambda_i - j,\,\,\,\,\,\,\, l_{\lambda}(s) =
\lambda^{t}_{j}-i\] are hook arm-length and leg-length of box
$s=(i,j)$ of the Young tableau respectively.

The structure of this article is as following. In section 2, we
review the refined topological vertex formulation in the A-model
setup and its applications to instanton counting problems of
$N=2^{\ast}$ theories. The eCS model and its spectrum which is
captured by Jack symmetric functions, are described in section 3. In
section 4, we show that the Jack symmetric functions exactly
reproduce the Nekrasov instanton partition function as expected.
This computation confirms the relation between topological string
theoty, which geometric engineers the $N=2^{\ast}$ theory, and the
2d eCS therory, which relates to the 4d theory by the AGT
relation\cite{Donagi,NekSha}. Section 5 is left for conclusions and
further interests.
\section{Refined Topological Vertex and instanton counting in $N=2^{\ast}$  theories}
The refined topological vertex(RTV) is a two-parameter
generalization of the ordinary topological vertex. In the
topological vertex formulation, one can easily get the partition
function of an A-model which generates an $N=2$ gauge theory by
geometric engineering. On the other hand, the same $N=2$ theory can
be obtained by the NS5-D4 brane setup of IIA string theory. The
bridge between these two apparently different configurations is the
large $n$ transition. On the field theory side, the nonperturbative
part of the partition function is captured by the Nekrasov instanton
counting, which involves the so-called $\Omega$ deformation of
$\mathbb R^4$. On the topological A-model side, the $\Omega$
deformation relates to the two-parameter generalization of the
topological vertex, which is the RTV. The refined partition function
of topological string is equivalent to the Nekrasov partition
function of $N=2$ theories\cite{RTV,Aganagic,Taki,Awata05,Awata09}.

Since we will frequently use the relation between these two
procedures, it is necessary to review the refined topological vertex
and its connection with Nekrasov's partition function.
\subsection{Brane setup and toric diagram}
The brane setup of $N=2$ theories can be translated into toric
diagrams of topological A-model as following. One draws the brane
intersection diagram of a desired $N=2$ theory as in Fig.3a, then
blows up every 4-vertex as two 3-vertices, adjusts the toric diagram
to match with the geometric engineering
procedure\cite{GE}\footnote{For a toric CY-3fold related to a gauge
theory, there should exists a preferred direction in which all
gluing legs of the toric diagram are parallel.}, as showing in
Fig.3b.

\begin{figure}
\centering
\includegraphics{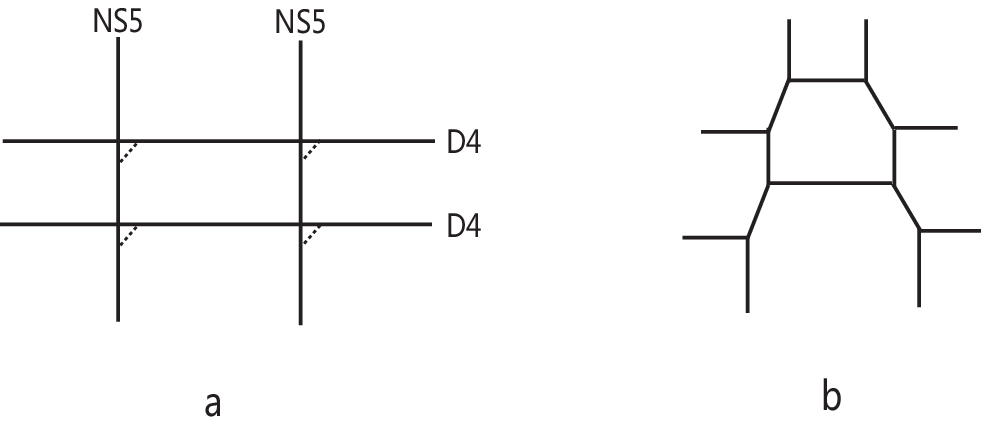}
\caption{\small{\emph{a. NS5-D4 configuration of $N=2\,\, U(2)$
theory with $N_f=4$,\,\,\, b. The related toric diagram}}}
\label{Fig3.eps}
\end{figure}

 From the NS5-D4 intersection branes configuration, one can
immediately read off its low energy effective theory is just the
$N=2$ gauge theory. The pure gauge part of the theory comes from the
coincided D4 branes, the matters are due to the truncation of the
two NS5 branes \footnote{Our main considerations in the present
article do not involve fundamental matters. In the brane setup they
do not only from the infinity D4 branes ending on the left or the
right of NS5 branes, but also can be alternatively realized as the
addition of D6 branes.}. However, the topological string realization
of the $N=2$ theory is totally different. The pure gauge part comes
from the blowup of the singularities of the ALE space in Calabi-Yau.
The matters correspond to D-branes wrapping on Lagrangian
submanifolds in Calabi-Yau.

The detailed relation of these two realizations of $N=2$ gauge
theories were considered in Dijkgraaf and Vafa's article
\cite{DV2009a} which we will now briefly review in the following.
Instead of the A-model, they considered the mirror B-model
realization. The Coulomb parameters of the gauge theory which are
positions of D4 branes in their transverse directions, relate to the
large $n$ limit of the condensation of D2 branes, or equivalently,
the condensation of the screen charges in the 2d CFT language of the
B-model. The matters are related to the insertions of stacks of D2
branes which can be written as vertex operators in 2d CFT. Their
masses correspond to the numbers of branes. The Nekrasov $\Omega$
deformation is translated to a phase changing of the complex
coordinate of the spectral curve. We will come back to these points
in the second part of this note.

\subsection{The refined topological vertex}
The refined topological vertex is defined as \cite{RTV}
\begin{eqnarray}
C_{\lambda\mu\nu}(t,q) &=&
\left(\frac{q}{t}\right)^{\frac{\parallel\mu\parallel^2+\parallel\nu\parallel^2}{2}}t^{\frac{\kappa}{\mu}}P_{\nu^t}(t^{-\rho};q,t)\\\nonumber&\times&
\sum_{\eta}\left(\frac{q}{t}\right)^{\frac{|\eta|+|\lambda|-|\mu|}{2}}s_{\lambda^t/\eta}(t^{-\rho}q^{-\nu})s_{\mu/\eta}(t^{-\nu^t}q^{-\rho})\\\nonumber
P_{\nu^t}(t^{-\rho};q,t) &=&
t^{\frac{\parallel\nu\parallel^2}{2}}\tilde{Z}_{\nu}(t,q) =
\prod_{s\in\nu}\left(1-t^{l_{\nu}(s)+1}q^{a_{\nu}(s)}\right)^{-1}\\\nonumber
t=e^{\beta\epsilon_1},\,\,\,\,q=e^{-\beta\epsilon_2},&&\parallel\mu\parallel^2
= \sum_i \mu_i^2,\,\,\,\,\,\rho =
\{-\frac{1}{2},-\frac{3}{2},-\frac{5}{2},\cdots\}
\end{eqnarray}
where $\lambda,\mu,\nu$ denote Young tableaus of partitions of
instantons. $s_{\lambda}$ and $s_{\lambda/\eta}$ are the Schur and
the skew Schur function which is briefly reviewed in Appendix.A.
$P_{\nu^t}(t^{-\rho};q,t)$ is the Macdonald function.
\begin{figure}
\centering
\includegraphics{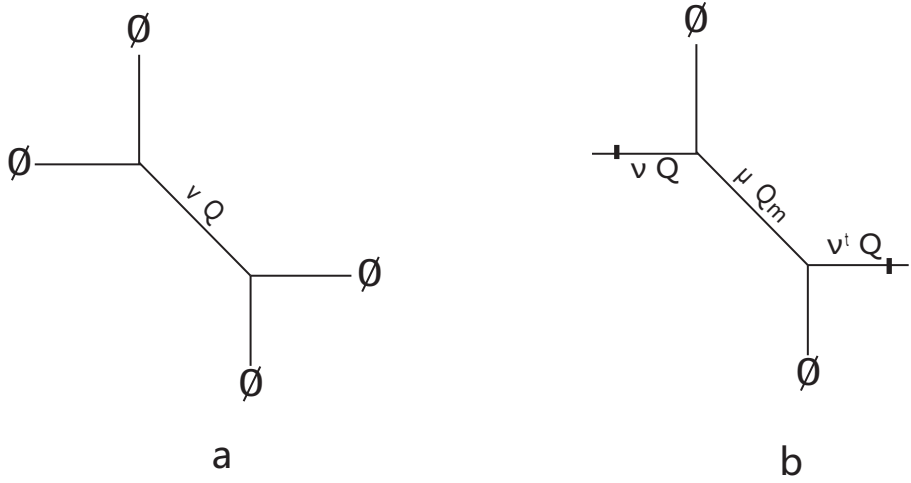}
\vspace{-40mm}\caption{\small{\emph{a. Toric diagram for ${\cal
O}(-1)\oplus{\cal O}(-1)\mapsto\mathbb{P}^1$,\,\,\,b. Toric diagram
of U(1) with a single adjoint matter}}} \label{Fig4.eps}
\end{figure}

For a toric diagram describing  a chosen CY-3fold, the refined
partition function can be calculated by gluing all topological
vertices.\footnote{If there are framing differences in the gluing
process, one should also introduce the framing factors. In this
article, we will assume all the edges in toric diagram are in the
standard framing.} For ${\cal O}(-1)\oplus{\cal
O}(-1)\mapsto\mathbb{P}^1$ as in Fig.\ref{Fig4.eps}a, the refined
partition function can be written as
\begin{eqnarray}
Z(t,q,Q)&=&\sum_{\nu}Q^{|\nu|}(-1)^{|\nu|}C_{{\o}{\o}\nu}(t,q)C_{{\o}{\o}\nu^t}(q,t)\\\nonumber
&=&
\sum_{\nu}Q^{|\nu|}(-1)^{\nu}q^{\frac{\parallel\nu\parallel^2}{2}}t^{\frac{\parallel\nu^t\parallel^2}{2}}\tilde{Z}_{\nu}(t,q)\tilde{Z}_{\nu^t}(q,t)\\\nonumber
&=&
\sum_{\nu}\frac{Q^{|\nu|}(-1)^{\nu}q^{\frac{\parallel\nu\parallel^2}{2}}t^{\frac{\parallel\nu^t\parallel^2}{2}}}
{\prod_{s\in\nu}(1-t^{l_{\nu}(s)+1}q^{a_{\nu}(s)})(1-t^{l_{\nu}(s)}q^{a_{\nu}(s)+1})}.
\end{eqnarray}

For more complicated toric diagram, the calculation principle is the
same.

\subsection{Refined partition functions for 5D
$N=2^{\ast}$ theories}
\subsubsection{U(1) theory}
 The simplest
5D $N=2^{\ast}$ theory is the $U(1)$ gauge theory with a single
adjoint hypermultiplet\cite{Iqbal, Aganagic}. The toric diagram
looks the same as the ${\cal O}(-1)\oplus{\cal
O}(-1)\mapsto\mathbb{P}^1$ but partially compactifying the two
external legs as shown in Fig.4b. Now the refined partition function
reads
\begin{eqnarray}\label{5DU(1)}
Z^{5D}_{\nu,\nu^t}(Q, Q_m, t, q) &=&
\sum_{\mu,\nu}(-Q)^{|\nu|}(-Q_m)^{|\mu|}C_{{\o}\mu\nu}(t,q)C_{\o\mu^t\nu^t}(q,t)\\\nonumber
&=&
\sum_{\lambda,\mu,\eta^1,\eta^2}(-Q)^{|\nu|}(-Q_m)^{|\mu|}\left(\frac{q}{t}\right)^{\frac{\parallel\mu\parallel^2+\parallel\nu\parallel^2}{2}}
\left(\frac{t}{q}\right)^{\frac{\parallel\mu^t\parallel^2+\parallel\nu^t\parallel^2}{2}}t^{\frac{\kappa(\mu)}{2}}q^{\frac{\kappa(\mu^t)}{2}}
\\\nonumber
 &\times&t^{\frac{\parallel\nu\parallel^2}{2}}q^{\frac{\parallel\nu^t\parallel^2}{2}}\tilde{Z}_{\nu}(t,q)\tilde{Z}_{\nu^t}(q,t)
 s_{\mu}(t^{-\nu^t}q^{-\rho})s_{\mu^t}(t^{-\rho}q^{-\nu})\\\nonumber
 &=&
 \sum_{\nu}(-Q)^{|\nu|}t^{\frac{\parallel\nu^t\parallel^2}{2}}q^{\frac{\parallel\nu\parallel^2}{2}}\frac{\prod_{i,j=1}^{\infty}(1-Q_m t^{-\nu^t_i-\rho_j}q^{-\nu_j-\rho_i})}
 {\prod_{s\in\nu}(1-t^{l_{\nu}(s)+1}q^{a_{\nu}(s)})(1-t^{l_{\nu}(s)}q^{a_{\nu}(s)+1})}.
\end{eqnarray}
This 5D refined partition function contains a perturbative part
which is just the zero-instanton part $Z^{5D}_{\o}(Q,Q_m,t,q)$, thus
the pure instanton part is given by
\begin{eqnarray}\label{U(1)inst}
Z^{5D}_{inst}(Q_m, t,q) &=& \frac{Z^{5D}_{\nu,\nu^t}(Q, Q_m, t,
q)}{Z^{5D}_{\o,\o}(Q,Q_m,t,q)}=\sum_{\nu}(-Q)^{|\nu|}\left(\frac{q}{t}\right)^{\frac{|\nu|}{2}}\\\nonumber
&\times&\prod_{(i,j)\in\nu}\frac{(1-Q_m
t^{-\nu^t_i-\rho_j}q^{-\nu_j-\rho_i})(1-Q_m
t^{\nu^t_i+\rho_j}q^{\nu_j+\rho_i})}{(1-t^{-l_{\nu}(s)-1}q^{-a_{\nu}(s)})(1-t^{l_{\nu}(s)}q^{a_{\nu}(s)+1})}
\end{eqnarray}
\subsubsection{U(2) theory} We now consider the U(2) theory using the same RTV
formulation. The toric diagram is showed in Fig. 6, the 5D refined
partition function is given by
\begin{figure}
\centering
\includegraphics{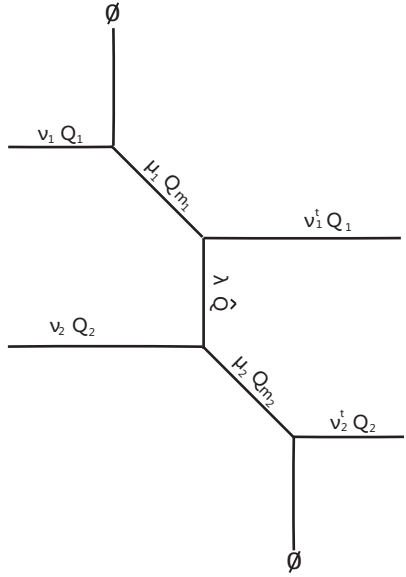}
\caption{\small{\emph{Toric diagram for U(2) N=2* theory}}}
\label{Fig5.eps}
\end{figure}
\begin{eqnarray}
Z^{5D}_{\nu_1,\nu_1^t; \nu_2, \nu_2^t}(U(2)) &=& \sum_{{\nu_i},
{\mu_i},
\lambda}\prod_{i=1}^2(-Q_{i})^{|\nu_i|}(-Q_{m_i})^{|\mu_i|}(-Q)^{|\lambda|}\\\nonumber
&\times&C_{\o\mu_1\nu_1}(t,q)C_{\lambda\mu_1^t\nu_1^t}(q,t)C_{\lambda^t\mu_2\nu_2}(t,q)C_{\o\mu_2^t\nu_2^t}(q,t)\\\nonumber
&=&\sum_{{\nu_i}, {\mu_i},{\eta_i}
\lambda}\prod_{i=1}^2(-Q_{i})^{|\nu_i|}(-Q_{m_i})^{|\mu_i|}(-Q)^{|\lambda|}\left(\frac{q}{t}\right)^{\frac{|\eta_1|-|\eta_2|}{2}}
t^{\frac{\parallel\nu_1^t\parallel^2+\parallel\nu_2^t\parallel^2}{2}}q^{\frac{\parallel\nu_1\parallel^2+\parallel\nu_2\parallel^2}{2}}\\\nonumber
&\times&\tilde{Z}_{\nu_1}(t,q)\tilde{Z}_{\nu_1^t}(q,t)\tilde{Z}_{\nu_2}(t,q)\tilde{Z}_{\nu_2^t}(q,t)s_{\mu_1^t/\eta_1}(t^{-\rho}q^{-\nu_1})s_{\mu_1}(q^{-\rho}t^{-\nu_1^t})\\\nonumber
&\times&
s_{\mu_2/\eta_2}(t^{-\nu_2^t}q^{-\rho}s_{\mu_2^t}(q^{-\nu_2}t^{-\rho})s_{\lambda^t/\eta_1}(t^{-nu_1^t}q^{-\rho})s_{\lambda/\eta_2}(t^{-\rho}q^{-\nu_2})\\\nonumber
&=&\sum_{\nu_1,\nu_2}t^{\frac{\parallel\nu_1^t\parallel^2+\parallel\nu_2^t\parallel^2}{2}}q^{\frac{\parallel\nu_1\parallel^2+\parallel\nu_2\parallel^2}{2}}
\tilde{Z}_{\nu_1}(t,q)\tilde{Z}_{\nu_1^t}(q,t)\tilde{Z}_{\nu_2}(t,q)\tilde{Z}_{\nu_2^t}(q,t)\\\nonumber
&\times& \prod_{i,j=1}^{\infty}\frac{1-Q
t^{\rho_i-\nu_{1,j}^t}q^{-\rho_j-\nu_{2,i}}(1-QQ_{m_1}Q_{m_2}
t^{-\rho_i - \nu_{1,j}^t}q^{-\rho_j-\nu_{2,i}})}{1-QQ_{m_1}
t^{i-1-\nu_{1,j}^t}q^{j-\nu_{2,i}}}\\\nonumber &\times&
\frac{(1-Q_{m_1} t^{-\rho_i-\nu_{1,j}^t}q^{-\rho_j -
\nu_{1,i}})(1-Q_{m_2} t^{-\rho_i-\nu_{2,j}^t}q^{-\rho_j -
\nu_{2,i}})}{(1-QQ_{m_2} t^{i-\nu_{1,j}^t} q^{j-1-\nu_{2,i}})}.
\end{eqnarray}
The instanton part of the refined partition reads
\begin{eqnarray}\label{U(2)inst}
Z_{inst}^{5D}(U(2)) &=& \frac{Z^{5D}_{\nu_1,\nu_1^t; \nu_2,
\nu_2^t}(U(2))}{Z^{5D}_{\o,\o;
\o,\o}(U(2))}=\sum_{\nu_1,\nu_2}\prod_{i=1}^2(-\sqrt{\frac{q}{t}}Q_{i})^{|\nu_i|}\\\nonumber
&\times& \prod_{(j,k)\in\nu_i}(1-Q_{m_i}
t^{-\rho_j-\nu_{i,k}^t}q^{-\rho_k - \nu_{i,j}})(1-Q_{m_i}
t^{\rho_j+\nu_{i,k}^t}q^{\rho_k + \nu_{i,j}})\\\nonumber &\times&
\prod_{(j,k)\in\nu_1}(1-Q_i'
t^{\rho_j+\nu_{2,k}^t}q^{\rho_k+\nu_{1,j}})\prod_{(j,k)\in\nu_2}(1-Q_i'
t^{-\rho_j-\nu_{1,k}^t}q^{-\rho_k-\nu_{2,j}})\\\nonumber &\times&
\left[\prod_{(j,k)\in\nu_1}(1-QQ_{m_1}
t^{-j+\nu_{2,k}^t}q^{\nu_{1,j}-k+1})\prod_{(j,k)\in\nu_2}(1-QQ_{m_1}
t^{j-\nu_{1,k}^t-1}q^{-\nu_{2,j}+k})\right]^{-1}\\\nonumber &\times&
\left[\prod_{(j,k)\in\nu_1}(1-QQ_{m_2}
t^{-j+\nu_{2,k}^t+1}q^{\nu_{1,j}-k})\prod_{(j,k)\in\nu_2}(1-QQ_{m_2}
t^{j-\nu_{1,k}^t}q^{-\nu_{2,j}+k-1})\right]^{-1}\\\nonumber &\times&
\left[\prod_{s\in\nu_i}(1-t^{-l_{\nu_i}(s)-1}q^{-a_{\nu_i}(s)})(1-t^{l_{\nu_i}(s)}q^{a_{\nu_i}(s)+1})\right]^{-1},
\end{eqnarray}
here we define $Q_1' = Q,\,\,\,\,\, Q_2' = QQ_{m_1}Q_{m_2}$.

\subsubsection{$U(2)\times U(2)$ theory}
The toric diagram for $N=2^{\ast}$ $U(2)\times U(2)$ theory is given
in Fig. 6a. Using the gluing rule\cite{Taki,Aganagic}, one can
truncate the toric diagram into two separate ones denoted by $T_1$
and $T_2$(as showing in Fig. 6b). The 5D refined partition for $T_1$
and $T_2$ are
\begin{figure}
\centering
\includegraphics[scale=0.8]{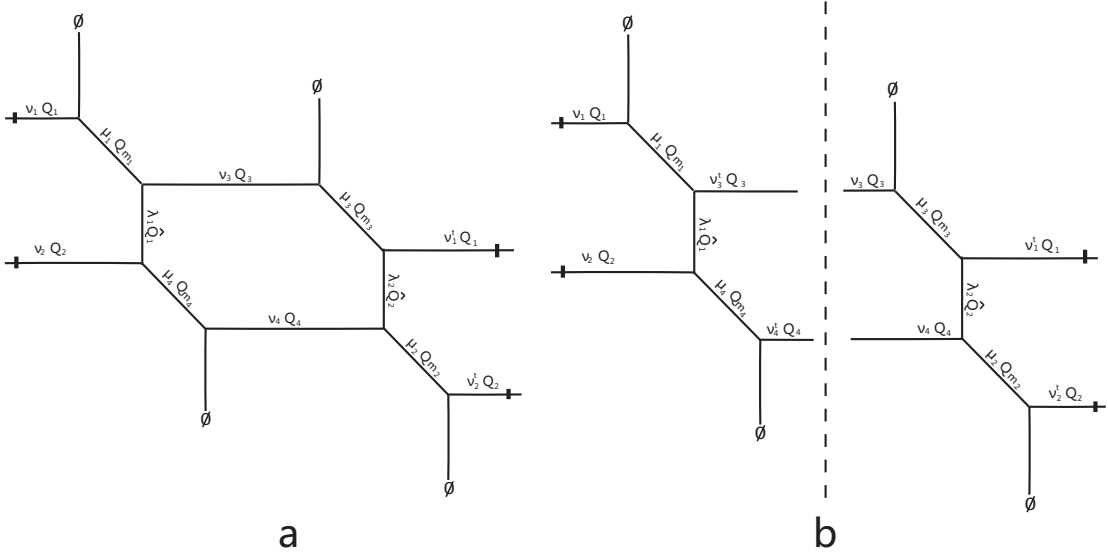}
\vspace{-30mm}\caption{\small{\emph{a. Toric diagram $U(2)\times
U(2)$,\,\,\,b. The gluing of toric diagrams $T_1$ and $T_2$}}}
\label{Fig6.eps}
\end{figure}
\begin{eqnarray}
Z_{\nu_1,\nu_3^t; \nu_2, \nu_4^t}^{T_1, 5D} &=&
Z^{5D}_{\nu_1,\nu_3^t; \nu_2, \nu_4^t}(U(2))
\\\nonumber &=&
(-Q_{m_1})^{|\mu_1|}(-Q_{m_4})^{|\mu_4|}(-\hat{Q}_1)^{|\lambda_1|}(-Q_{1})^{|\nu_1|}(-Q_{2})^{|\nu_2|}\\\nonumber
&\times&
C_{\o\mu_1\nu_1}(t,q)C_{\lambda_1\mu_1^t\nu_3^t}(q,t)C_{\lambda_1^t\mu_4\nu_2}(t,q)C_{\o\mu_4^t\nu_4^t}(q,t),\\
Z_{\nu_3,\nu_1^t; \nu_4, \nu_2^t}^{T_2, 5D} &=&
Z^{5D}_{\nu_3,\nu_1^t; \nu_4, \nu_2^t}(U(2))
\\\nonumber &=&
(-Q_{m_2})^{|\mu_2|}(-Q_{m_3})^{|\mu_3|}(-\hat{Q}_2)^{|\lambda_2|}(-Q_{3})^{|\nu_3|}(-Q_{4})^{|\nu_4|}\\\nonumber
&\times&
C_{\o\mu_3\nu_3}(t,q)C_{\lambda_2\mu_3^t\nu_1^t}(q,t)C_{\lambda_2^t\mu_2\nu_4}(t,q)C_{\o\mu_2^t\nu_2^t}(q,t),
\end{eqnarray}
respectively. The instanton part is given by
\begin{eqnarray}
Z^{5D}_{inst}(U(2)\times U(2)) = \frac{Z_{\nu_1,\nu_3^t; \nu_2,
\nu_4^t}^{T_1, 5D}Z_{\nu_3,\nu_1^t; \nu_4, \nu_2^t}^{T_2,
5D}}{Z_{\o,\o; \o, \o}^{T_1, 5D}Z_{\o,\o; \o, \o}^{T_2, 5D}}.
\end{eqnarray}
After an elementary calculation, we get:
\begin{eqnarray}\label{5DU(2)2}
Z_{inst}^{5D} &=&
\sum_{\{\nu_i\}}\prod_{i=1}^4\left(-\sqrt{\frac{q}{t}}Q_{i}\right)^{|\nu_i|}\\\nonumber
&\times& \prod_{\{r,s\}}\prod_{(j,k)\in\nu_r}(1-Q_{m_s}
t^{-\rho_k-\nu_{s,j}^t}q^{-\rho_j-\nu_{r,k}})(1-Q_{m_r}
t^{\rho_k+\nu_{s,j}^t}q^{\rho_j+\nu_{r,k}})\\\nonumber
&\times&\prod_{m=1}^2\prod_{(j,k)\in\nu_2}(1-\hat{Q}_{1,m}t^{-\rho_j-\nu_{3,k}^t}q^{-\rho}_k-\nu_{2,j})
\prod_{(j,k)\in \nu_3}(1-\hat{Q}_{1,m}t^{\rho_j+\nu_{2,k}^t
q^{\rho_k +\nu_{3,j}}})\\\nonumber
&\times&\prod_{n=1}^2\prod_{(j,k)\in\nu_4}(1-\hat{Q}_{2,n}t^{-\rho_j-\nu_{1,k}^t}q^{-\rho}_k-\nu_{4,j})
\prod_{(j,k)\in \nu_1}(1-\hat{Q}_{2,n}t^{\rho_j+\nu_{4,k}^t
q^{\rho_k +\nu_{1,j}}})\\\nonumber
&\times&\left[\prod_{s\in_{\nu_i}}(1-t^{-l_{\nu_i}(s)-1}q^{-a_{\nu_i}(s)})(1-t^{l_{\nu_i}(s)}q^{a_{\nu_i}(s)+1})\right]^{-1}
\\\nonumber
&\times&\left[\prod_{\{p,q\}}\prod_{(j,k)\in\nu_p}(1-\tilde{Q}_p
t^{j-\nu_{q,k}^t}q^{-\nu_{p,j}+k-1})\prod_{(j,k)\in\nu_q}(1-\tilde{Q}_p
t^{\nu_{p,k}^t -j+1} q^{\nu_{q,j}-k}q^{-\nu_{p,j}+k-1})\right]^{-1},
\end{eqnarray}
where
\begin{eqnarray}
\nonumber\{r,s\}\in\{1,3\} \,\,or \,\,\{2,4\}, r\neq
s,\,\,\,\,\{p,q\}\in\{1,2\} \,\,or \,\,\{3,4\}, p\neq q\\\nonumber
\hat{Q}_{1,1} = Q_{m_1},\,\,\,\hat{Q}_{1,2} =
\hat{Q}_2Q_{m_2}Q_{m_3},\,\,\,\hat{Q}_{2,1} =
Q_{m_2},\,\,\,\hat{Q}_{2,2} = \hat{Q}_1Q_{m_1}Q_{m_4}\\\nonumber
\tilde{Q}_1 = \hat{Q}_1Q_{m_1},\,\,\tilde{Q}_2 =
\hat{Q}_2Q_{m_2}\frac{q}{t},\,\,\tilde{Q}_3 =
\hat{Q}_2Q_{m_3},\,\,\tilde{Q}_4 = \hat{Q}_2Q_{m_4}\frac{q}{t}.
\end{eqnarray}
If one defines the following identites
\begin{eqnarray}
\nonumber
Z_{\nu^{(\ell)}_a,\nu^{(\ell+1)}_b}^{bifund,\,\,5D}(Q^{(\ell,\ell+1)}_{ab},t,q)&=&\prod_{(i,j)\in\nu_a^{(\ell)}}
(1-Q^{(\ell,\ell+1)}_{ab}t^{\nu_{b,j}^{(\ell+1)t}-i}q^{\nu_{a,i}^{(\ell)}-j+1})\\\nonumber&\times&\prod_{(i,j)\in\nu_a^{(\ell+1)}}
(1-Q^{(\ell,\ell+1)}_{ab}t^{-\nu_{a,j}^{(\ell)t} +i
-1}q^{-\nu_{b,i}^{(\ell+1)}+j})\\\nonumber Z^{vec,
\,\,5D}_{\nu^{(\ell)}_a,\nu^{(\ell)}_b}(Q^{\ell}_{ab},t,q) &=&
\left[\prod_{(i,j)\in\nu_a}(1-Q^{(\ell)}_{ab}t^{\nu_{b,j}^t-i}q^{\nu_{a,i}-j+1})\prod_{(i,j)\in\nu_b}(1-Q^{(\ell)}_{ab}t^{-\nu_{a,j}^t
+i -1}q^{-\nu_{b,i}+j})\right]^{-1}\\
 Z^{vec,
\,\,5D}_{\nu^{(\ell)}_a,\nu^{(\ell)}_a}(t,q)
&=&\left[\prod_{(i,j)\in\nu_a}(1-t^{-l_{\nu_a}(s)-1}q^{-a_{\nu_a}(s)})(1-t^{l_{\nu_a}(s)}q^{a_{\nu_a}(s)+1})\right]^{-1},
\end{eqnarray}
the above refined 5D instanton partition can be written as
\begin{eqnarray}
Z_{inst}^{5D} &=&
\sum_{\{\nu_i\}}\prod_{i=1}^4\left(-\sqrt{\frac{q}{t}}Q_{m_i}\right)^{|\nu_i|}
\prod_{a,b
=1}^2Z_{\nu^{(\ell)}_a,\nu^{(\ell+1)}_b}^{bifund,\,\,5D}(Q^{(\ell,\ell+1)}_{ab},t,q)\\\nonumber
&\times&Z^{vec,
\,\,5D}_{\nu^{(\ell)}_a,\nu^{(\ell)}_b}(Q^{(\ell)}_{ab},t,q)Z^{vec,
\,\,5D}_{\nu^{(\ell)}_a,\nu^{(\ell)}_a}(t,q).
\end{eqnarray}
This formula coincides with Nekrasov's instanton partition function
for $5D\,\,\, N=2\,\,\, U(2)\times U(2)$ gauge theory.

\subsubsection{$U(N)$ M-node necklace quiver theory}
The generalization to $U(N)$ M-node necklace quiver theory (Fig.9)
is straightforward, and the result has the following expression
\begin{figure}
\centering
\includegraphics[scale=1.2]{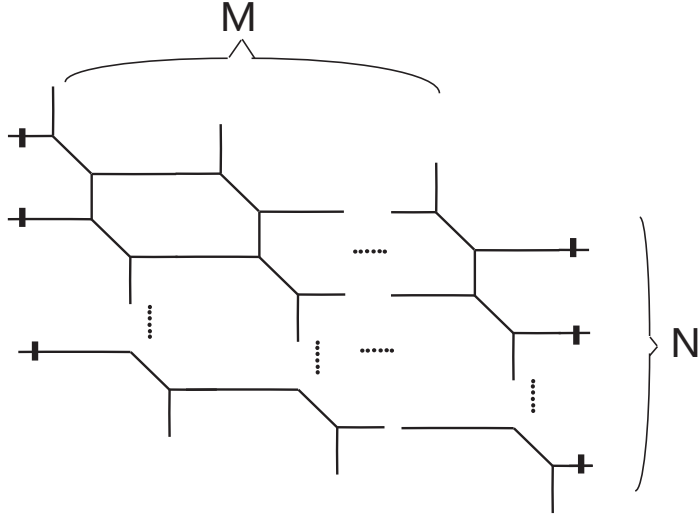}
\caption{\small{\emph{Toric diagram for elliptic M-node $U(N)$
theory}}} \label{Fig7.eps}
\end{figure}
\begin{eqnarray}
Z^{U(N)\,\,\,inst}_{M-necklace} &=&
\prod_{i=1,\ell=1}^{N,M}(-\sqrt{\frac{q}{t}}Q^{(\ell)}_{i})^{\nu^{(\ell)}_i}\prod_{a,b
=1}^N
Z_{\nu^{(\ell)}_a,\nu^{(\ell+1)}_b}^{bifund,\,\,5D}(Q^{(\ell,\ell+1)}_{ab},t,q)\\\nonumber
&\times&Z^{vec,
\,\,5D}_{\nu^{(\ell)}_a,\nu^{(\ell)}_b}(Q^{(\ell)}_{ab},t,q)Z^{vec,
\,\,5D}_{\nu^{(\ell)}_a,\nu^{(\ell)}_a}(t,q).
\end{eqnarray}
This can be easily proved by using the RTV formulation and the
mathematical induction.
\subsection{4D field theory limit}
To compare the refined partition functions with the real 4D
theories, one could shrink the perimeter of the cyclic 5-th
dimension to zero, that is, $\beta\rightarrow 0$.

For $U(1)$ theory, the parameters are set as
\[Q_m = \sqrt{\frac{t}{q}}e^{\beta (-\tilde{m})} = e^{\beta({\epsilon_+}/2-\tilde{m})},\,\,\, Q = \sqrt{\frac{q}{t}}e^{\beta(-\tilde{a}+\tilde{m})} =
e^{\beta(-\tilde{a}+\tilde{m}-\epsilon_+/2)},\] where
\[m=\frac{\tilde{m}}{\epsilon_2},\,\,a=\frac{\tilde{a}}{\epsilon_2}.\]
Then the U(1) instanton partition function reads
\begin{eqnarray}
{\bf Z}^{4D, U(1)}_{inst}&=&\text{lim}_{\beta\rightarrow
0}{}Z^{5D}_{inst}(Q_m, t,q)\\\nonumber &=&
\sum_{\nu}(\sqrt{\frac{q}{t}}Q)^{|\nu|}\prod_{s\in\nu}\frac{(-m+\beta
l(s)+a(s)+1)(-m-\beta (l(s)+1)-a(s))}{(\beta(a(s)+1)+l(s))(\beta
a(s)+1+l(s))}
\end{eqnarray}

For $U(2)$ theory, the parameters are set as
\[Q_{m_1}=Q_{m_2}=\sqrt{\frac{t}{q}}e^{\beta (-\tilde{m})},\,\,\,\, QQ_{m_1}=QQ_{m_2}=e^{\beta(-\tilde{a})}.\]
The instanton partition function for this theory is
\begin{eqnarray}
\nonumber{\bf Z}^{4D,U(2)}_{inst} &=& \text{lim}_{\theta\rightarrow
0}Z_{inst}^{5D}(U(2))\\\nonumber &=&
\sum_{\nu_1,\nu_2}\prod_{i=1,j=1}^2(\sqrt{\frac{q}{t}}Q_i)^{|\nu_i|}\\\nonumber&\times&\prod_{s\in\nu_i}(-m_{i,j}+a_{\nu_i}(s)+1+\beta
l_{\nu_j}(s))\prod_{t\in\nu_j}(-m_{i,j}-a_{\nu_j}(t)-\beta
(l_{\nu_i}(t)+1))\\\nonumber &\times& \left[\prod_{i\neq
j}\prod_{s\in\nu_i}(a_{i,j}+a_{\nu_i}(s)+1+\beta
l_{\nu_j}(s))\prod_{t\in\nu_j}(a_{i,j}-a_{\nu_j}(t)-\beta
(l_{\nu_i}(t)+1))\right]^{-1}\\\nonumber&\times&\left[\prod_{s\in\nu_i}(\beta(l_{\nu_i}(s)+1)+a_{\nu_i}(s))(\beta
l_{\nu_i}(s)+1+a_{\nu_i}(s))\right]^{-1},
\end{eqnarray}
here\[m_{i,j} = a_i-a_j-m, \,\,\,a_{i,j}=a_i-a_j,
\,\,\,a_{1,2}=-a_{2,1}=a.\] Now one can immediately read off the
expression (\ref{M-Necklace}) we proposed in the introduction
section. This is just a substitution \[\tilde{Q}_{i} =
\sqrt{\frac{q}{t}}Q_i, \,\,\,\, \lambda_{\ell,i} = \nu^{(\ell)}_i.\]

The $U(2)\times U(2)$ instanton partition function is just a product
of two $U(2)$ ones,
\begin{eqnarray}
\nonumber{\bf Z}^{4D,U(2)\times U(2)}_{inst} &=&
\text{lim}_{\theta\rightarrow 0}Z_{inst}^{5D}(U(2)\times
U(2))\\\nonumber
&=&\sum_{\vec{\nu}_1,\vec{\nu}_2}\prod_{i,\ell=1}^{2}(
\tilde{Q}_{i})^{|\vec{\nu}_i|}\prod_{j,k=1}^{2}{\Large\mbox
{$[$}}(\langle E^{a^{(i)}_{j,k}}(E^{\ast})^{\beta - a^{(i)}_{j,k}-1}
J_{\nu_{i,j}}, J_{\nu_{i,k}}\rangle_{\beta}|_{j\neq k})\langle
J_{\nu_{i,j}}, J_{\nu_{i,j}}\rangle_{\beta}{\Large\mbox
{$]$}}^{-1}\nonumber\\&&\times \prod_{m,n=1}^{2}\langle
E^{m^{(\ell,\ell+1)}_{m,n}}(E^{\ast})^{\beta -
m^{(\ell,\ell+1)}_{m,n}-1} J_{\nu_{\ell,m}},
J_{\nu_{\ell+1,n}}\rangle_{\beta}.
\end{eqnarray}
It is easy to generalize to the M-node quiver $U(N)$ theory. The
result is given in Eq.(\ref{M-Necklace}).
\section{Jack symmetric functions and eCS models}
Since the $N=2^{\ast}$ theories are all superconformal field
theories, according to the 4d-2d relation proposed by Alday, Gaiotto
and Tachikawa \cite{AGT}, there are 2D Liouville/Toda integrable
systems corresponding to these gauge theories. The $N=2^{\ast}$
theories are related to the elliptic Calogero-Sutherland(eCS)
models. The CS model\footnote{The eCS model can be seen as the
analytic continuation of the original CS model.} plays an important
role in many subjects in physics and mathematics. Such as conformal
field theory(CFT), unitary matrix models, fractional quantum hall
effects(FQHE), etc. Its spectrum can be totally released from the
so-called Jack polynomials. In principle, the total system can be
solved by using the properties of Jack polynomials, from the CFT
point of view, Jack polynomials have natural meanings of characters
of the symmetry which drives the model. For instance, the Jack
polynomials related to certain Young tableaux are believed to
correspond to the singular vectors of $W$-algebra, this algebra
reflects the hidden $W_{1+\infty}$ symmetry of CS model.

The instanton counting of the $N=2^{\ast}$ theories should be
related to the counting of the BPS spectrums in 4D gauge theories.
In 2D point of view, this can be seen as the counting of the
admissible representations of the eCS model, that is, the counting
of singular vectors in the model. As pointed out in Awata,
Sakamoto's works\cite{Awata95, Sakamoto} on singular vectors in CS
model, Jack polynomials and skew Jack polynomials define the
singular vector space under $W$-algebra\footnote{Actually, the Jack
polynomials associated with rectangular Young tableaux are singular
vectors of Virasoro algebra. The non-rectangular ones are related to
W-algebra.}.

\subsection{Jack polynomials and Calogero-Sutherland model}
The Hamiltonian of Calogero-Sutherland model is given by:
\begin{equation}
H=P_i^2+\beta(\beta-1) sin^{-2} (\frac{1}{2}(x_i-x_j)) \\
=(-i\partial_i-iA_i)(-i\partial_i+iA_i) \\
=-\partial_i^2+A_i^2-\sum _i \partial_i A_i
\end{equation}
here  \(\partial_i=\partial_{x_i}\),  \(A_i=\sum_i \beta ctg
x_{ij}\) its ground state captures by the equation of motion:
\[(-i\partial_i+iA_i)\psi_0 =0,\] the general solution gives
\[\psi_0=\sum_{i<j}sin^{\beta}(x_i-x_j).\]
Define the excitation state as \(\psi_\lambda=J_{\lambda}\psi_0\),
thus it should satisfy \[[-\partial_i^2+\beta \sum ctg x_{ij}
(\partial_i -
\partial_j)] J_{\lambda} = c_{\lambda}J_{\lambda}.\]
It is not hard to get the operator formulism for this
$\hat{J}_{\lambda}$. However, there are very simple vertex operator
maps from Calogero-Sutherland model to CFT. Denote
\begin{gather}
\psi_0= \langle k_f|V_k(z_1)\cdots V_k(z_n)|k_i\rangle \\\nonumber
z_i=e^{ix_i}\\\nonumber V_k (z_i)= e^{ik\phi(z_i)}\\\nonumber
\end{gather}
and choose choose proper vacuum momentum \(k_i\) as that
\begin{equation}
\psi_0 \sim \prod_{i<j}^N (z_i-z_j)^{k^2}\prod_{i=1}^N z_i^{k_i
\cdot k}\nonumber,
\end{equation}
then the excitation state is just as following
\begin{eqnarray}\label{CSjack}
\psi_{\lambda}&=& \langle k_f|\hat{J_{\lambda}}V_k(z_1)\cdots
V_k(z_n)|k_i\rangle\\\nonumber  \hat{J}_{\lambda}&=&\sum_n
d_{\lambda}^{[n]} \hat{P}_{[n]}=\sum_n d_{\lambda}^{[n]} \frac
{\hat{a}_{[n]}}{(\sqrt{\beta}) ^{l(\lambda)}}\\\nonumber
\hat{a}_{[n]}&=&\hat{a}_{n_1} \cdots \hat{a}_{n_l},\nonumber
\end{eqnarray}
here $\hat{P}_{[n]}$ is the operator formulism of Newton polynomial,
$\ell(\lambda)$ is the total number of rows in $\lambda$.
$d_{\lambda}^{[n]}$is the normalization factor such that the
normalization of $J_{\lambda}(z^i)$(for the partition
$\lambda=\{j^{k_j}\}$)
\begin{eqnarray}
\langle J_{\lambda}, J_{\mu}\rangle_{\theta} &=&
\delta_{\lambda\mu}d_{\lambda}^{[n]}d_{\mu}^{[n]}\langle
k_f|\frac{\hat{a}_{\lambda}\hat{a}_{-\mu}}{\beta^{\frac{1}{2}\ell(\lambda)}\beta^{\frac{1}{2}\ell(\mu)}}|k_i+Nk\rangle\\\nonumber
&=&z_{\vec{\lambda}_j}\beta^{-\ell(\lambda)}d_{\lambda}^{[n]}d_{\mu}^{[n]}
= \delta_{\lambda\mu}j_{\lambda},\\\nonumber
j_{\lambda}&=&\prod_{s\in\lambda}(a_{\lambda}(s)+\beta(l_{\lambda}(s)+1))(\beta
l_{\lambda}(s) +a_{\lambda}(s)+1),
\\\nonumber z_{\vec{\lambda}_j}&=& \prod_{j=1}^{\infty}j^{k_j}k_j!\,\,\,\,,
\end{eqnarray}here \(k_i \longrightarrow k_i +Nk\) reflects the action of the zero
modes of vertex operators. By the mode expansion of the free boson
field $\phi(z)$, one reaches
\begin{gather}
\nonumber\phi(z)=\hat{q}+\hat{p}lnz +\sum_{n\in
\mathbb{Z},n\neq0}\frac {\hat{a}_{-n}}{n} z^n\\\nonumber V_k(z)=e^{k
\cdot \phi(z)}.
\end{gather}
Substitute these to Eq.(\ref{CSjack}), it is easy to show that
\begin{eqnarray}
\nonumber\psi_{\lambda}&=& J_{\lambda} \psi_0 \\\nonumber &=&\langle
k_f| d_{\lambda}^{[n]} \frac {\hat{a}_{[n]}}{(\sqrt{\beta})
^{l(\lambda)}} \prod_{i<j}^N (z_i-z_j)^{k^2}\prod_{i=1}^N z_i^{k_i
\cdot k} e^{\sum_{m \in \mathbb{Z}^{+}}k \frac {\hat{a}_{-m}}{m}
z_i^m}|k_i+Nk\rangle
\\\nonumber&=&\langle k_f| d_{\lambda}^{[n]} \frac {\sum_i k
z_i^{n_1}}{\sqrt{\beta}} \frac {\sum_i k z_i^{n_2}}{\sqrt{\beta}}
\cdots \frac {\sum_i k z_i^{n_l}}{\sqrt{\beta}} \psi_0({z_i})
|k_i+Nk\rangle
\end{eqnarray}
 \(e^{\sum_{m \in \mathbb{Z}^{+}}k \frac
{\hat{a}_{-m}}{m} z_i^m}\)is the remaining \(\prod_i^N
V_k^{+}\)after normal ordering, we have used the relation \(\hat{a}
e^{\hat{a}^{+} \alpha} |0\rangle= \alpha  e^{\hat{a}^{+} \alpha }
|0\rangle\) from the second step to the third step of the above
expression. If \(k=\sqrt{\beta}\), we see that the Jack polynomial
$J_{\lambda}(z)$ is actually can be seen as the excitation state of
the CS model.
\subsection{Screening charges and singular vectors}
It is shown in Dijkgraaf and Vafa's article\cite{DV2009a} the
screening charges are related to instanton insertions. The screening
charges of CS model are defined as in\cite{Awata95, WXYjack}
\[\alpha_+ = k, \,\,\,\alpha_- = -\frac{1}{k},\]
then by the Felder's cohomology\cite{Felder} and using Thorn's
method\cite{Awata95, Thorn,  WXYjack}, one can easily prove that the
singular vector $\ket{\chi_{-r,-s}^+}$ associated with rectangular
Young tableau $\lambda = \{s^r\}$
 can be written as
\begin{eqnarray}
  \ket{\chi_{-r,-s}^+}
  &\!\!=\!\!&
  \oint\prod_{j=1}^r\frac{dz_j}{2\pi i}\cdot
  \prod_{i=1}^r:e^{\alpha_+\phi(z_i)}:
  \ket{\alpha_{r,-s}} \\\nonumber
  &\!\!=\!\!&
  \oint\prod_{j=1}^r\frac{dz_j}{2\pi iz_j}\cdot
  \prod_{i,j=1 \atop i<j}^r(z_i-z_j)^{2\beta}\cdot
  \prod_{i=1}^rz_i^{(1-r)\beta-s}\cdot
  \prod_{j=1}^re^{\alpha_+\phi_-(z_j)}
  \ket{\alpha_{-r,-s}}\\\nonumber
  &\!\!\!\!&
\ket{\alpha}=e^{\alpha\hat{q}}\ket{0},\,\,\,\alpha_{r,s}=\frac{(1-r)\alpha_+}{2}+\frac{(1-s)\alpha_-}{2}
\end{eqnarray}
the integratation contours have been chosen as the Felder's contours
as in Fig.8

{~}\vspace{5mm}
%%%%%%%%%%%%%%%%%%%%%%%%%%%%%%%%%%%%
%                                  %
%      Figure 2                    %
%                                  %
%%%%%%%%%%%%%%%%%%%%%%%%%%%%%%%%%%%%
%\input epsf.tex
\centerline{\includegraphics[bb=0 0 2.5in 2.5in]{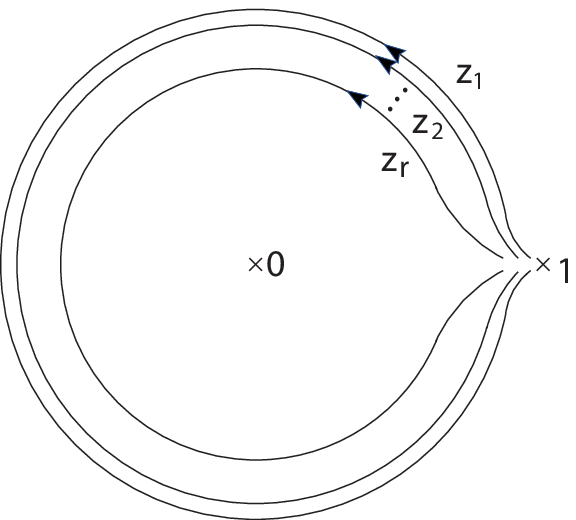}}
\centerline{{\bf Figure. 8:} Felder's Integration contours}
%%%%%%%%%%%%%%%%%%%%%%%%%%%%%%%%%%%%%
{~}\vspace{3mm} The Jack polynomial can be identified with the
following expression \begin{eqnarray}
  {\cal N}_{r,s}^+ \,{\cal N}_{(s^r)}^+ J_{(s^r)}(x)
  &\!\!=\!\!&
  \bra{\alpha_{r,s}}C_{k}\ket{\chi_{r,s}^+} \\\nonumber
  &\!\!=\!\!&
  \oint\prod_{j=1}^r\frac{dz_j}{2\pi iz_j}\cdot
  \prod_{i,j=1 \atop i<j}^r(z_i-z_j)^{2\beta}\cdot
  \prod_{i=1}^rz_i^{(1-r)\beta-s}\cdot
  \prod_i\prod_{j=1}^r(1-w_iz_j)^{-\beta}\\\nonumber
  &&C_{k}=e^{k\sum_{n>0}\frac{1}{n}a_n p_n} =
  \prod_{i}V^-_{k}(w_i),\,\,\,V^-_{k}(w_i)=e^{-k\phi_-(w_i)}
  \label{J+rs}
\end{eqnarray} where the normalization constants ${\cal N}_\lambda^+$ \cite{Stanley}
and ${\cal N}_{r,s}^+$ \cite{Awata95} are given by
\begin{eqnarray}
  {\cal N}_{\lambda}^+
  =
  \prod_{s\in\lambda}
  \frac{(\ell_{\lambda}(s)+1)\beta+a_{\lambda}(s)}
       {\ell_{\lambda}(s)\beta+a_{\lambda}(s)+1}, \qquad
  {\cal N}_{r,s}^+
  =
  \frac{1}{r!}
  \prod_{j=1}^r\frac{\sin\pi j\beta}{\sin\pi\beta}\cdot
  \frac{\Gamma(r\beta+1)}{\Gamma(\beta+1)^r}.
  \label{Nrs}
\end{eqnarray}
Similarly, as proved in \cite{Awata95}, the Jack polynomials
associated with non-rectangular Young tableaux are related to the
singular vectors for $W_N$-algebra.
\begin{eqnarray}
 \ket{\chi_{\vec{r},\vec{s}}^-}
  =
  \oint\prod_{a=1}^{N-1}\prod_{j=1}^{r^a}\frac{dz^a_j}{2\pi i}\cdot
  \prod_{a=1}^{N-1}\prod_{j=1}^{s^a}:e^{\alpha_+\phi^a(z^a_j)}:
  \ket{\vec{\lambda}_{\vec{r},\vec{s}}^-
  -\alpha_+\sum_{a=1}^{N-1}r^a\vec{\alpha}^a}
\end{eqnarray}
with $s^1>\cdots>s^{N-1}$. The corresponding Young tableau is showed
as follows. \generalYoung The operator formalism of generic Jack
polynomial can be identified with the insertion between the left and
the right vacuum denote by $\bra{\lambda_{\vec{r},\vec{s}}}$ and
$\ket{\vec{\lambda}_{\vec{r},\vec{s}}^-
  -\alpha_+\sum_{a=1}^{N-1}r^a\vec{\alpha}^a}$. It follows
\begin{eqnarray}
\hat{J}_{\lambda}\sim\prod_{i}V^-_{k}(w_i)\oint\prod_{a=1}^{N-1}\prod_{j=1}^{r^a}\frac{dz^a_j}{2\pi
i}\cdot
  \prod_{a=1}^{N-1}\prod_{j=1}^{s^a}:e^{\alpha_+\phi^a(z^a_j)}:
\end{eqnarray}
\section{Jack polynomial and Nekrasov's instanton partition function}

 The existence of singular vectors implies that correlation
functions in CS model can be split into conformal blocks. These
conformal blocks, due to the AGT relation, should be exact the
instanton partition function of the related $N=2^{\ast}$ theory.
However, for the M-node necklace quiver gauge theory, the related 2D
correlation function is still hard to calculate. However, we can
read off that there should have a more simple description of this
correlation function from the result we obtained in present paper.
It is just simple multiplications of two point functions within the
insertion of two Jack polynomials! This is a factorization formulism
rather than a summation, the combinatorial properties of conformal
blocks are totally determined by the Jack polynomials. Now we
extract these information at the level of result. We will explain
the hidden physics using Dijkgraaf-Vafa's mirror B-model picture in
the second part of this note.

The deformation parameters of the eCS model can be written as \bea
\epsilon_1=ig_sk,\,\, \epsilon_2=
\frac{ig_s}{k},\,\,Q=k+\frac{1}{k}=\frac{\epsilon_1+\epsilon_2}{ig_s}.\,\,\eea

The bifundamental part  is the building block of the instanton
partition function. The expression reads \bea{\bf
Z}_{bifund}^{4D\,\,\,inst}(\vec{a}_{\ell},\vec{\lambda}_{\ell},\vec{a}_{\ell
+1},\vec{\lambda}_{\ell +1}; m_{\ell}) = \prod_{m,n=1}^{N}\langle
E^{m^{(\ell,\ell+1)}_{m,n}}(E^{\ast})^{\beta -
m^{(\ell,\ell+1)}_{m,n}-1} J_{\lambda_{\ell,m}},
J_{\lambda_{\ell+1,n}}\rangle_{\beta}.\eea The insersion of
$E^{m^{(\ell,\ell+1)}_{m,n}},\,\,m^{(\ell,\ell+1)}_{m,n} =
a_{\ell+1,n}-a_{\ell,m}-m_{\ell}$ can be rewritten as follows
\begin{eqnarray}
\bra{0}C_k\text{exp}{(\frac{\tilde{m}}{-\epsilon_2}\frac{(-1)^np_n}{n})}
&=&\bra{0}C_k\text{exp}{(\frac{im}{g_s}\frac{(-1)^na_n}{n})}\\\nonumber
&=&\bra{0}C_k\prod_{\tilde{m}}\Gamma^-(-1) \\\nonumber
&=&\bra{0}s_{\mu}(-1,-1,-1,\cdots)C_k
\\\nonumber
&=&\sum_{\mu}\bra{\mu}(-1)^{|\mu|}C_k.
\end{eqnarray}
The conjugate state induced by the insertion of
$(E^{\ast})^{\beta-m-1}$ has the same express except the conjugate
charge is given by $\epsilon_+ - m$ as expectation. The whole
expression now is given by
\begin{eqnarray}
\sum_{\mu}\bra{\mu}(-1)^{|\mu|}\hat{J_{\lambda}}\hat{J_{\nu}}\sum_{\mu'}(-1)^{|\mu'|}\ket{\mu'}\\\nonumber
\end{eqnarray}
When one expands the $s_{\mu}$ as the monomial symmetric function
$m_{\mu}$, and writes the Jack polynomial as the complete
homogeneous symmetric function $h_{\lambda}$\footnote{This can be
done as that in Stanley's article\cite{Stanley} and Macdonald's
textbook\cite{Macdonald}.}, one immediately gets the right
expression of the inner product. The insertions of $E$ and
$E^{\ast}$ have a explanation that $m$ Wilson loops translated
between the associated branes, also, this will be shown in the
second part of this note.

\section{Conclusions and discussions}

We calculate in this note the instanton partition function of the
elliptic N=2 M-node quiver gauge theory using the refined
topological vertex formulation. The result exactly coincident with
Nekrasov's instanton partition. We find the instanton counting of
$N=2^{\ast}$ theories has a neat expression in terms of Jack
polynomials as expected\cite{NekSha}. We give a explanation of the
expression at the level of result. This result implies that the AGT
duality between 4D $N=2$ supersymmetric gauge theories and the 2D
conformal field theories has more refined structures such as the
physical reason of the factorization of conformal blocks.
\section*{Acknowledgement}
The author thanks Yingying Xu and Song He for useful discussions on
topological vertex and instanton counting. Prof. Ming Yu gave a
great many of supports on the operator formalism of Jack
polynomials.

\end{document}